\documentclass[reprint,showpacs,aps,superscriptaddress,pra]{revtex4-1}
\usepackage{graphics}
\usepackage{bm}
\usepackage{bbm}
\usepackage{amsmath}
\usepackage{amssymb}
\usepackage{graphicx}
\usepackage{amsthm}
\begin{document}

\title{Fermionic Entanglement in Superconducting Systems}
\author{M.\ Di Tullio} 
\affiliation{IFLP/CONICET and Departamento  de F\'{\i}sica,
    Universidad Nacional de La Plata, C.C. 67, La Plata (1900), Argentina}
\author{N.\ Gigena}
\affiliation{IFLP/CONICET and Departamento  de F\'{\i}sica,
    Universidad Nacional de La Plata, C.C. 67, La Plata (1900), Argentina}
\author{R.\ Rossignoli}
\affiliation{IFLP/CONICET and Departamento  de F\'{\i}sica,
    Universidad Nacional de La Plata, C.C. 67, La Plata (1900), Argentina}
\affiliation{Comisi\'on de Investigaciones Cient\'{\i}ficas (CIC), La Plata (1900), Argentina}

\begin{abstract}
We examine distinct measures of fermionic entanglement in the exact ground
state of a finite superconducting system. It is first shown that global measures such
as the one-body entanglement entropy, which represents the minimum relative entropy between the exact ground state 
and the set of fermionic gaussian states, exhibit a close correlation with the BCS
gap, saturating in the strong superconducting regime. The same behavior is displayed by the bipartite 
entanglement between the set of all single particle states $k$ of positive quasimomenta  and their 
 time reversed partners $\bar{k}$. In contrast, the entanglement associated with the reduced 
 density matrix of four single particle modes $k,\bar{k}$, $k',\bar{k}'$, which can be measured through a 
 properly defined fermionic concurrence, exhibits a different behavior, showing a peak in the vicinity of the
superconducting transition for states  $k,k'$ close to the fermi level and 
becoming small in the strong coupling regime. In the latter such reduced state
exhibits, instead, a finite mutual information and quantum discord. And while  the 
first measures can be correctly estimated with the BCS
approximation, the previous four-level concurrence lies strictly beyond the
latter, requiring at least a particle number projected BCS treatment for its
description. Formal properties of all previous entanglement measures are as
well discussed.

\end{abstract}
\maketitle

\section{Introduction}

Quantum entanglement is an essential feature of quantum mechanics. The basic
notion was developed for systems of distinguishable components
\cite{NC.00,Sc.95,We.89,BD.96}, where it has been extensively investigated
\cite{A.08,H.09,E.10}, playing a key role in fundamental quantum information
processing tasks \cite{BB.93,JL.03}. On the other hand, the theory of
entanglement for systems of indistinguishable components such as fermions, is
more recent
\cite{Sch.01,SDM.01,Eck.02,Wi.03,GM.04,Ci.09,Za.02,Shi.03,Fri.13,Be.14,Pu.14,
Ci.09,Pl.09,IV.13,Os.14,Os.142,SL.14,IV.14,GR.15,M.16,GR.16,DB.16,GR.17,AF.17}, and
is subject to some debate \cite{KCP.14}. There are  essentially two distinct
approaches. One is based on the entanglement between modes
\cite{Za.02,Shi.03,Fri.13,Be.14,Pu.14,GR.17}. Here the subsystems are defined
in terms of subsets of single particle modes, and entanglement depends
obviously on the choice of basis for the single particle state space and then on the choice
of modes for each subsystem. The other, known as entanglement between particles
\cite{Sch.01,SDM.01,Eck.02,Wi.03,GM.04,Ci.09,Ci.09,Pl.09,IV.13,Os.14,Os.142,SL.14,IV.14}
considers the indistinguishable constituents as subsystems and entanglement is
defined beyond antisymmetrization, such that a Slater determinant is not
entangled.

In \cite{GR.15} we defined a one-body entanglement entropy for fermion systems,
which for pure states is determined by the one-body density matrix and vanishes
if and only if (iff) the state is a Slater determinant.  It can be obtained from a single particle mode 
entanglement measure after optimization over all possible bases of the single particle 
space. The approach can be directly extended to deal with states with no fixed
fermion number (though still  having fixed number parity \cite{Fr.15}), in
which case it vanishes iff the state is a quasiparticle vacuum.  In the case of
a single particle space of dimension $4$, the approach is an extension of the entanglement
measure developed in \cite{Sch.01} for pure states with fixed fermion number,
and provides a lower bound to the entanglement associated with any  bipartition
of the single particle space \cite{GR.17}. In such space its convex roof extension can also
be analytically evaluated for any mixed state in terms of the fermionic
concurrence \cite{Sch.01,GR.15}.

The aim of this work is to analyze  the previous general measures of fermionic
entanglement in the exact ground state of a finite superconducting system.
Previous studies of entanglement in such systems focused mainly on the formal
properties of pairing correlations \cite{Ci.09}  or on the Bardeen-Cooper-Schrieffer (BCS) state
\cite{Pu.14,DLZ.05,OK.05,CL.08},  using in this case methods and measures
specifically devised for such state. Here we first show that the one-body
entanglement entropy is in the present system a direct indicator of pairing
correlations, reflecting  essentially the BCS gap and saturating in the strong
superconducting regime. It is also shown to be closely related to the bipartite
mode entanglement between the $\Omega$ states $k$ and their time-reversed
partners, becoming strictly proportional to it at the BCS level. On the other
hand, the fermionic entanglement associated with four single particle modes $k\bar{k}$,
$k'\bar{k}'$, exhibits a different behavior. This quantity is determined
by a mixed reduced state with no fixed fermion number yet fixed (even) number
parity,  and  can be explicitly evaluated through the fermionic concurrence
defined in \cite{GR.15}. It shows a peak in the vicinity of the ground state superconducting
transition for levels $k,k'$ close to the Fermi level, becoming then small in
the strong superconducting regime (if the system size is not too small). In the
latter, this reduced state exhibits instead classical and discord-type
\cite{OZ.01,HV.03,Mo.11,AB.16} correlations, leading to a finite mutual
information and quantum discord. We also discuss the BCS description of these
quantities, showing that it can indeed provide a correct estimation of the
first measures in the superconducting phase, although it fails to describe the
four-mode fermionic concurrence, which is identically zero in BCS for all
coupling strengths. This quantity is shown to require at least a projected BCS
treatment.

In sec.\  \ref{II} we first discuss the main  properties of the employed
fermionic entanglement measures, some of them not included in
\cite{GR.15,GR.17}, showing in particular their direct relation with the
minimum relative entropy to a fermionic gaussian state. It is also shown that
in the case of four single particle modes, their extension to mixed states also warrants, if
non-zero,  a finite bipartite mode entanglement for any partition of the single particle 
space. The application of these measures to the exact ground state of a finite
superconducting system is discussed in sec.\ \ref{IIIA}, where their behavior
as a function of the pairing coupling strength is analyzed. Their description
through the BCS approximation is discussed in \ref{IIIB}, which also includes a
simple projected (before variation) BCS treatment, necessary for describing the
four-mode fermionic concurrence. Other quantities like the mutual information
and quantum discord of  four single particle modes, are also discussed in \ref{III} and in
the Appendices, which contain additional details. Exact analytic  expressions
for the strong coupling regime are as well provided.

 \section{Formalism}
 \label{II}
\subsection{One-body entanglement entropies}
We consider a fermion system described by a single particle  space ${\cal H}$,
spanned by fermion operators $c_i, c^\dagger_i$, $i=1,\ldots,d$,
satisfying the anticommutation relations $\{c_i,c^\dagger_j\}=\delta_{ij}$,
$\{c_i,c_j\}=\{c^\dagger_i,c^\dagger_j\}=0$.
Given a pure state $|\Psi\rangle$ of this system, the set of averages
\begin{equation}\rho^{\rm sp}_{ij}=\langle c^\dagger_j c_i\rangle\equiv\langle\Psi|
c^\dagger_j c_i|\Psi\rangle\label{rsp}\,,\end{equation} form the one-body
density matrix $\rho^{\rm sp}=\mathbbm{1}-\langle \bm{c}\bm{c}^\dagger\rangle$
($\bm{c}=(c_1,\ldots,c_d)^t$). It plays the role of a ``reduced'' density
matrix  which determines the average of any one-body operator
$\hat{O}=\bm{c}^\dagger O\bm{c}=\sum_{i,j}O_{ij}c^\dagger_i c_j$:
$\langle\hat{O}\rangle={\rm tr}\,[\rho^{\rm sp}O]$, with ${\rm tr}$ denoting
the trace in the single particle  space.

In \cite{GR.15} we have defined an associated one-body  entanglement entropy,
\begin{eqnarray}
E(|\Psi\rangle)&=&{\rm tr}\,[h(\rho^{\rm sp})]=
\sum_i h(f_i) \,,\label{1}\\
h(f_i)&=&-f_i\log_2 f_i-(1-f_i)\log_2 (1-f_i)\,,
\end{eqnarray}
where $f_i=\langle a^\dagger_i a_i\rangle$ are the eigenvalues of $\rho^{\rm sp}$ and  
$\bm{a}=U^\dagger\bm{c}$ is the set of fermion operators diagonalizing $\rho^{\rm sp}$, such
that $\langle a^\dagger_j a_i\rangle=(U^\dagger \rho^{\rm
sp}U)_{ij}=f_{i}\delta_{ij}$, with $U^\dagger U=\mathbbm{1}$.

Eq.\ (\ref{1}) vanishes iff $f_i=0$ or $1$ $\forall\, i$, i.e., iff
$|\Psi\rangle$ is a Slater determinant ($|\Psi\rangle=[\prod_i (a^\dagger_i)^{f_i}]|0\rangle$),
and remains obviously invariant under one-body unitary transformations
$|\Psi\rangle\rightarrow \exp[-i\bm{c}^\dagger O\bm{c}] |\Psi\rangle$
($O^\dagger=O$), which just lead to a unitary transformation of $\rho^{\rm sp}$ 
($\rho^{\rm sp}\rightarrow e^{-iO}\rho^{\rm sp} e^{iO}$). It is also the
minimum, over all single particle bases, of the entropy determined by the average occupation
of the corresponding single particle states  \cite{GR.15}:
\begin{equation}
E(|\Psi\rangle)=\mathop{\rm Min}_{\{c_i\}}\sum_i h(\langle c^\dagger_i c_i\rangle)\,,\label{min}
\end{equation}
with  $h(\langle c^\dagger_i c_i\rangle)$ representing the
entanglement entropy of a single particle mode with the remaining modes (see also sec.\ \ref{12}).

Eq.\ (\ref{1}) also admits other interpretations. It can be regarded as the von
Neumann entropy $S(\rho')$, in the grand canonical ensemble, of the
independent fermion density operator $\rho'$ which reproduces the whole single
particle density matrix determined by $|\Psi\rangle$: If
\begin{eqnarray}
\rho'&=&Z^{-1}\exp[-\bm{c}^\dagger\Lambda \bm{c}]
=Z^{-1}\exp[-\sum_i \lambda_i a^\dagger_i a_i]\,,\label{rhop}
\end{eqnarray}
with $Z={\rm Tr}\exp[-\bm{c}^\dagger\Lambda\bm{c}]=\prod_i(1+e^{-\lambda_i})$
and $\lambda_i$ the eigenvalues of the matrix $\Lambda$, then
\begin{eqnarray}
S(\rho')=-{\rm Tr}\,[\rho'\log_2\rho']=\sum_i h(f_i)\,,\label{srhop}
\end{eqnarray}
where $f_i={\rm Tr}[\rho' a^\dagger_i a_i]=[1+e^{\lambda_i}]^{-1}$. Eq.\ (\ref{srhop}) 
will then coincide with (\ref{1})
provided these  $f_i$'s  are  identical with the eigenvalues of the single particle density matrix  
(\ref{rsp}), i.e., provided
 \begin{equation}
\mathbbm{1}-{\rm tr}[\rho'\bm{c}\bm{c}^\dagger]=[\mathbbm{1}+\exp(\Lambda)]^{-1}=\rho^{\rm sp}\,,
\label{rsp2}
\end{equation}
which implies $\Lambda=\ln[(\rho^{\rm sp})^{-1}-\mathbbm{1}]$.

This result shows that Eq.\ (\ref{1})  is in fact the  {\it relative entropy} \cite{Wh.78,Ve.02}
(in the grand canonical ensemble) between the pure state  $\rho=|\Psi\rangle\langle\Psi|$
and the state (\ref{rhop}) which satisfies (\ref{rsp2}), since $S(\rho)=0$ and
(\ref{rsp2}) implies ${\rm Tr}[\rho \log_2 \rho']={\rm Tr}[\rho'\log_2\rho']$:
\begin{eqnarray}
S(\rho||\rho')&\equiv&-{\rm Tr}\,[\rho(\log_2\rho'-\log_2\rho)]\label{srel00}\\
&=&S(\rho')={\rm tr}\,[h(\rho^{\rm sp})] \,.\label{srel0}
\end{eqnarray}
Moreover, as shown in Appendix \ref{A}, Eq.\ (\ref{srel0}) is also the {\it
minimum}  relative entropy (in the grand canonical ensemble)  between $\rho$ and {\it any}
operator of the form (\ref{rhop}):
\begin{equation}
\mathop{\rm Min}_{\rho'}S(\rho||\rho')={\rm tr}\,[h(\rho^{\rm sp})]\,.\label{min2}
\end{equation}
Hence, Eq.\ (\ref{1}) is a measure of the minimum distance between $\rho$ and
the set of operators of the form (\ref{rhop}) (fermionic gaussian states
commuting with $N$).

{\it Extension to quasiparticles}. If the state  $|\Psi\rangle$ does not have a
fixed fermion number $N=\sum_i c^\dagger_i c_i$, (but has a definite number
parity $e^{i\pi N}=\pm 1$, in agreement with the parity superselection
rule \cite{Fr.15}), we can define a generalized one-body entanglement entropy
\cite{GR.15} based on the extended  one-body density matrix $\rho^{\rm qsp}$,
which contains the contractions $\kappa_{ij}=\langle c_j c_i\rangle$ and
$-\kappa_{ij}^*=\langle c^\dagger_jc^\dagger_i\rangle$:
\begin{eqnarray}
&&E^{\rm qsp}(|\Psi\rangle)=-{\rm tr}[\rho^{\rm qsp}\log_2\rho^{\rm qsp}]=
\sum_i h(\tilde{f}_i)\,,\label{Eq}\\
&&\rho^{\rm qsp}=\mathbbm{1}-\left\langle\begin{pmatrix}
\bm{c}\\\bm{c}^{\dagger\,t}\end{pmatrix}\begin{pmatrix}
\bm{c}^\dagger&\bm{c}^t\end{pmatrix}\right\rangle=
\begin{pmatrix}\rho^{\rm sp}&\kappa\\-\kappa^*&{\mathbbm 1}-(\rho^{\rm sp})^*\end{pmatrix}
\,.\;\;\label{rqsp}
\end{eqnarray}
Here $\tilde{f}_i=\langle \tilde{a}^\dagger_i \tilde{a}_i\rangle$ and
$1-\tilde{f}_i=\langle \tilde{a}_i \tilde{a}^\dagger_i\rangle$ are the
eigenvalues of $\rho^{\rm qsp}$ (which always come in pairs
$(\tilde{f}_i,1-\tilde{f}_i)$), with $\tilde{a}_i$ the fermion quasiparticle
operators diagonalizing $\rho^{\rm qsp}$, related to the original operators
$c_i$, $c^\dagger_i$ through a Bogoliubov transformation \cite{RS.80}. Eq.\
(\ref{Eq}) reduces to (\ref{1}) iff $\kappa=0$, and vanishes iff $|\Psi\rangle$
is a Slater determinant or also a quasiparticle vacuum (or equivalently, a quasiparticle Slater determinant,
which can be always written as a quasiparticle vacuum through a  particle-hole
transformation). Eq.\ (\ref{Eq}) remains invariant under {\it arbitrary}
particle hole transformations  ($c_i\rightarrow c^\dagger_i$ for some single particle states
$i$), as well as {\it arbitrary Bogoliubov transformations}  \cite{GR.15}. It
is the minimum, over all single quasiparticle bases, of the sum of the
entanglement entropies of all single quasiparticle modes \cite{GR.15}.

Eq.\ (\ref{Eq}) is also the minimum relative entropy between $\rho$ and any
fermionic gaussian state, i.e..\ any $\rho'$  which is the exponent of a
generalized one-body operator:
\begin{eqnarray}
\mathop{\rm Min}_{\rho'}S(\rho||\rho')&=&-{\rm Tr}[\rho^{\rm qsp}\log_2 \rho^{\rm qsp}]\,,
\label{minqsp}\end{eqnarray}
\begin{eqnarray}
\rho'&=&Z^{-1}\exp[-\bm{c}^\dagger{\Lambda} \bm{c}-\frac{1}{2}(\bm{c}^\dagger\Gamma\bm{c}^{\dagger\,t}+
\bm{c}^t\Gamma^\dag\bm{c})]\label{rhoqsp}\\
&=&\tilde{Z}^{-1}\exp\left[-(\bm{c}^\dagger\;
\bm{c}^t){\cal L}{\small \begin{pmatrix}\bm{c}\\\bm{c}^{\dagger\,t}\end{pmatrix}}\right],
\;{\cal L}= \begin{pmatrix}
{\Lambda}&\Gamma\\-\Gamma^*&{\mathbbm 1}-{\Lambda}^*
\end{pmatrix}\,.\nonumber
\end{eqnarray}
The minimum (\ref{minqsp}) is reached for that $\rho'$ which reproduces the full $\rho^{\rm qsp}$, 
i.e., that satisfying
\begin{eqnarray}&&\mathbbm{1}-{\rm tr}\left[\rho' {\small\begin{pmatrix}\bm{c}\\\bm{c}^{\dagger\,t}
	\end{pmatrix}}(\bm{c}^\dagger\;
    \bm{c}^t)\right]=[1+\exp({\cal L})]^{-1}=\rho^{\rm qsp}\,,\nonumber\end{eqnarray}
which implies  ${\cal L}=\ln[(\rho^{\rm qsp})^{-1}-1]$ and hence
$S(\rho')=-{\rm tr}[\rho^{\rm qsp}\log_2 \rho^{\rm qsp}]$.

\subsection{Entanglement of bipartitions of the single particle space\label{12}}
Given a decomposition ${\cal H}_A\oplus{\cal H}_B$  of the single particle space ${\cal H}$
in orthogonal subspaces of finite dimension $d_A$, $d_B=d-d_A$, we may expand
$|\Psi\rangle$ in a set of Slater determinant in ${\cal H}_A$ and ${\cal H}_B$ as
$|\Psi\rangle=\sum_{\mu,\nu}\alpha_{\mu\nu}|\mu\nu\rangle$, where
$|\mu\nu\rangle=[\prod_{i\in A}(c^\dagger_i)^{n^\nu_i}][\prod_{j\in
B}(c^\dagger_j)^{n^\mu_j}]|0\rangle$, with $n^\nu_i=0,1$ the occupation number
of single particle state $i$ in Slater determinant $\nu$. The reduced states associated with these single particle 
subspaces are \cite{GR.16} $\rho_A=\sum_{\mu,\mu'}
(\alpha\alpha^\dagger)_{\mu\mu'}|\mu\rangle\langle\mu'|$ and
$\rho_B=\sum_{\nu,\nu'} (\alpha^t\alpha^*)_{\nu\nu'}|\nu\rangle\langle\nu'|$,
which  reproduce all expectation values of operators containing  creation and
annihilation operators acting just on these subspaces. They are normalized
mixed states with the same non-zero eigenvalues $\lambda_k$,  given by the
singular values of the matrix $\alpha$. Their entropies
$S(\rho_A)=S(\rho_B)=-\sum_k\lambda_k \log_2 \lambda_k$ represent the
entanglement entropy $E(A,B)$ associated with this partition
\cite{GR.16,GR.15}.

For states $|\Psi\rangle$ having definite fermion number, $\rho_{A(B)}$  will
commute with the local fermion number $N_{A(B)}=\sum_{i\in A(B)}c^\dagger_i
c_i$, but will in general be a mixture of states with different particle number
(it will be represented by a $2^{d_{A(B)}}\times 2^{d_{A(B)}}$ matrix).
Similarly, if $|\Psi\rangle$ has definite number parity,  $\rho_{A(B)}$ will commute with
the local number parity  $e^{i\pi N_{A(B)}}$, being a mixture of even and odd states.

For instance, if ${\cal H}_A$  involves just one single particle level $i$, $\rho_{A}\equiv
\rho_i$ will be determined by the average occupation $\langle c^\dagger_i
c_i\rangle$:
\begin{equation}
\rho_i=\begin{pmatrix}\langle c^\dagger_ic_i\rangle&0\\0&\langle c_ic^\dagger_i\rangle\end{pmatrix}\,,
\label{ri}\end{equation}
in the basis  $\{c^\dagger_i|0\rangle, |0\rangle\}$, with $\langle
c_ic^\dagger_i\rangle=1-\langle c_ic^\dagger_i\rangle$ ($\langle c_i\rangle=0$
due to  number parity  conservation). Its entropy $S(\rho_i)=h(\langle c^\dagger_i
c_i\rangle)$ represents the entanglement entropy of such mode with the
remaining modes. A single particle basis where  $S(\rho_i)=0$ $\forall$ $i$ exists iff  Eq.\
(\ref{1}) vanishes. And a single quasiparticle basis with the same property
exists iff $E^{\rm qsp}(|\Psi\rangle)=0$.

Similarly, if ${\cal H}_A$ comprises a pair of levels $i\neq j$, then
\begin{equation}
\rho_{ij}=\begin{pmatrix}\langle c^\dagger_i c_ic^\dagger_jc_j\rangle&0&0&\langle c_j c_i\rangle\\
0&\langle c^\dagger_ic_ic_jc^\dagger_j\rangle&\langle c^\dagger_j c_i\rangle&0\\0&\langle c^\dagger_i c_j\rangle&
\langle c_ic^\dagger_ic^\dagger_j c_j\rangle&0\\\langle c^\dagger_ic^\dagger_j\rangle &
0&0&\langle c_ic^\dagger_i c_jc^\dagger_j\rangle\end{pmatrix}\,,\label{rij}
\end{equation}
in the basis $\{c^\dagger_ic^\dagger_j|0\rangle, c^\dagger_i|0\rangle,
c^\dagger_j|0\rangle,|0\rangle\}$. This reduced state (${\rm Tr}\,\rho_{ij}=1$)
determines the average of any operator involving just $c_i,c_j,c^\dagger_i,
c^\dagger_j$, and its entropy $S(\rho_{ij})$  is the entanglement entropy  of
this   pair of single particle modes with the remaining modes. The outer (inner) block of
$\rho_{ij}$ corresponds to positive (negative) pair  number parity. In contrast with
$\rho_i$, $\rho_{ij}$ is not fully determined by $\rho^{\rm qsp}$ since its
diagonal elements (i.e.\ the probabilities of finding none, one or both levels
occupied) involve two-body contractions. Nonetheless, if $c_i,c_j$ are
operators diagonalizing $\rho^{\rm qsp}$ ($c_{i(j)}\rightarrow
\tilde{a}_{i(j)}$), $\rho_{ij}$ is obviously diagonal and $E(|\Psi\rangle)=0$
or $E^{\rm qsp}(|\Psi\rangle)=0$ implies $S(\rho_{ij})=0$ for such operators
(just one of the diagonal elements will be non-zero).

Note also that $S(\rho_{ij})$ depends on the subspace spanned by the single particle levels
$i,j$, but not on the particular choice of states $i,j$  within this subspace:
Any unitary or Bogoliubov transformation involving just $c_i, c^\dagger_i,c_j,
c^\dagger_j$ will leave such entropy invariant.

\subsection{Entanglement of subspaces of the single particle space}

Let us now consider the entanglement of a reduced state $\rho_A$ of a single particle 
subspace ${\cal H}_A$. For reduced states which commute with the local fermion
number $N_A$, we  define the associated one-body entanglement of formation as
\begin{equation}E(\rho_A)=\mathop{\rm Min}_{\sum_\alpha q_\alpha |\Psi^A_\alpha\rangle
	\langle\Psi^A_{\alpha}|=\rho_A}
q_\alpha E(|\Psi^A_{\alpha}\rangle)\,,\label{Ef}\end{equation}
where $q_\alpha\geq 0$,  $\sum_{\alpha} q_\alpha=1$ and the minimization 
is over all representations of $\rho_A$ as a convex
combination of pure states in ${\cal H}_A$  with definite fermion number.  
Eq.\ (\ref{Ef})  vanishes iff $\rho_A$ can be written
as a convex mixture of Slater Determinants.
Similarly, we define the generalized one-body entanglement of formation as
\begin{equation}E^{\rm qsp}(\rho_A)=\mathop{\rm Min}_{\sum_\alpha q_\alpha 
	|\Psi^A_\alpha\rangle\langle\Psi^A_{\alpha}|=\rho_A}
q_\alpha E^{\rm qsp}(|\Psi^A_{\alpha}\rangle)\,,\label{Ef2}\end{equation}
where the minimization is now over all representations of $\rho_A$ as  
convex combination of pure states with definite number parity.
Eq.\ (\ref{Ef2})  vanishes iff $\rho_A$ can be written as a convex mixture 
of quasiparticle vacua or Slater Determinants.

It is first apparent that if the full  state $|\Psi\rangle$ is a quasiparticle
vacuum or Slater determinant, then $E^{\rm qsp}(\rho_{A})=0$ {\it for  any subspace ${\cal
H}_A$:}  In this case all averages involved in the construction of $\rho_{A}$
can be determined through Wick's theorem \cite{RS.80}, and hence expressed in
terms of the elements of $\rho^{\rm qsp}$ involving this subspace. Therefore,
$\rho_{A}$ can be written as the exponent of a suitable generalized one-body
operator of the form (\ref{rhoqsp}) providing the same $\rho^{\rm qsp}$ for
this subspace, and will then be a convex combination of quasiparticle Slater Determinants or
vacua. A non-zero  value of  $E^{\rm qsp}(\rho_A)$ is then indicative of
correlations beyond those provided by a global quasiparticle vacuum. 
Similarly, if $|\Psi\rangle$ is a standard Slater determinant then  $E(\rho_{A})=0$, since in
this case $\kappa=0$ and $\rho_{A}$  can then be written as an operator of the
form (\ref{rhop}).

It is also apparent that for a single level $i$,  $E(\rho_i)=E^{\rm
qsp}(\rho_i)=0$ for any $|\Psi\rangle$. Similarly, for two single particle  levels we always
have $E^{\rm qsp}(\rho_{ij})=0$, since any pure state with fixed  number parity in a
two-dimensional single particle space (such as the eigenstates of $\rho_{ij}$) can be
written  as a quasiparticle vacuum or Slater determinant \cite{GR.15,GR.16}. And if $\langle
c_i c_j\rangle=0$ (i.e.\ $[\rho_{ij},N_{ij}]=0$) then $E(\rho_{ij})=0$. The
same property holds for three distinct levels $i,j,k$ for the same reason: Any
pure state with fixed  number parity in a three-dimensional single particle space can be written  as a
quasiparticle vacuum or Slater determinant \cite{GR.15}, implying $E^{\rm qsp}(\rho_{ijk})=0$
(and also  $E(\rho_{ijk})=0$ if $[\rho_{ijk},N_{ijk}]=0$).

The first non-trivial case is that of four distinct single particle levels $i,j,k,l$, in
which case a closed expression for the one-body entanglement of formation for
any state $\rho_{ijkl}$ with {\it fixed}  number parity was obtained in \cite{GR.15},
extending the results of \cite{Sch.01} for states with fixed fermion number.
The result is
\begin{equation}
E^{\rm qsp}(\rho_{ijkl})=-4\sum_{\nu=\pm} f_\nu \log f_\nu,
\;\;f_{\pm}={\textstyle\frac{1\pm\sqrt{1-C^2(\rho_{ijkl})}}{2}}\,,
\label{Ef3}
\end{equation}
where $C$ is the corresponding {\it fermionic concurrence} \cite{GR.15,Sch.01},
\begin{equation}C(\rho_{ijkl})=
{\rm Max}[2\lambda_{\rm max}-{\rm Tr}\,R(\rho_{ijkl}),0]\,,\label{Cf}\end{equation}
with $\lambda_{\rm max}$ the largest eigenvalue of
$R(\rho)=\sqrt{\rho^{1/2}\tilde{\rho}\rho^{1/2}}$ and  $\tilde{\rho}=T\rho^*T$
in a standard basis. The operation  $T$ is given explicitly in Appendix \ref{B}
(note that $\rho_{ijkl}$ and $T$ are represented by $8\times 8$ matrices).

For a pure $\rho_{ijkl}=|\Psi\rangle\langle\Psi|$, $C^2$ becomes a quadratic
entropy of the corresponding four-level  $\rho^{\rm qsp}$:
$C^2(|\Psi\rangle)=\frac{1}{2} {\rm Tr}\rho^{\rm qsp}(1-\rho^{\rm
qsp})=4f_+f_-$, with $f_{+}$,  $f_-=1-f_+$ the (four-fold degenerate)
eigenvalues of $\rho^{\rm qsp}$ \cite{GR.15}. For a mixed $\rho_{ijkl}$, the
result (\ref{Cf}) coincides with the convex roof extension of $C(|\Psi\rangle)$
[Eq.\ (\ref{Ef2}) with $E^{\rm qsp}(|\Psi^A_\alpha\rangle)$ replaced by
$C(|\Psi^A_\alpha\rangle)$]. If $\rho_{ijkl}$ commutes with  number parity but contains
components of both parities, it can be written as a mixture $\sum_{p=\pm} q_p
\rho_{ijkl}^p$ of two mixed states with definite  number parity and $E^{\rm
qsp}(\rho_{ijkl})$ can be evaluated as the average of the expressions for each
parity \cite{GR.15}.

We can also consider bipartitions $(A_1,A_2)$ of the ${\cal H}_A$ subspace,
with ${\cal A}_1$ and ${\cal A}_2$ of nonzero dimension and determined by a
given choice of levels in some single particle basis of ${\cal H}_A$,  and examine the
associated bipartite entanglement in the state  $\rho_{A_1,A_2}$. Such state
will be separable  iff it can be written as a convex combination of pure
product states (with definite  number parity) in $A_1$ and $A_2$, and entangled otherwise.
While there is in general no relation between this entanglement and $E^{\rm
qsp}(\rho_A$), in the case of four single particle levels, it was shown in \cite{GR.17} that
for a {\it pure} $\rho_{ijkl}=|\Psi\rangle\langle\Psi|$, the quantity
$\frac{1}{4}E^{\rm qsp}(|\Psi\rangle)$ provides a {\it lower bound} to the
entanglement entropy of {\it any} bipartition  of the single particle space.

For a general $\rho_{ijkl}$, we now show the following Lemma, which relates the
fermionic concurrence (\ref{Ef3}) with the entanglement of a bipartition: {\it
For a general four-level fermionic  state $\rho_{ijkl}$ commuting
with number parity and satisfying $C(\rho_{ijkl})> 0$,  {\it any}
bipartition of ${\cal H}_{ijkl}$ (like $(ij-kl)$ or $i-jkl$)  is  entangled}.\\
{\it Proof}: If $\rho_{ijkl}$ were separable for a given bipartition $A_1-A_2$,
it could  be written as a convex combination of pure product states
$|\mu_{A_1}\rangle\langle \mu_{A_1}|\otimes|\nu_{A_2}\rangle\langle
\nu_{A_2}|$, with $|\mu_{A_1}\rangle$, $|\nu_{A_2}\rangle$ having definite  number parity.
But since they are definite  number parity pure states in a single particle space of dimension $d\leq
3$, they are necessarily a Slater determinant or quasiparticle vacuum, as discussed above.
Consequently, $\rho_{ijkl}$ can be written as a convex combination of Slater determinant or
quasiparticle vacua  $|\mu_{A1}\nu_{A2}\rangle\langle \mu_{A_1}\nu_{A2}|$, of
definite  number parity, entailing $C(\rho_{ijkl})=0$. Thus, $C(\rho_{ijkl})>0$ ensures
that any bipartition of ${\cal H}_A$ is entangled, for {\it any} choice of single particle 
or quasiparticle basis of this subspace.\qed

\section{Application to a finite pairing system}
\label{III}
\subsection{Exact results \label{IIIA}}
We now consider a fermion system with a single particle space ${\cal H}$ of finite dimension
$2\Omega$,  spanned by  $\Omega$ orthogonal single particle states $k$ and the corresponding $\Omega$ time-reversed
states $\bar{k}$. We consider in such space a pairing Hamiltonian of
the form
\begin{eqnarray}
H&=&\sum_k \varepsilon_k(c^\dagger_k c_k+c^\dagger_{\bar{k}}c_{\bar{k}})-
\sum_{k, k'}G_{kk'}c^\dagger_{k'}c^\dagger_{\bar{k}'}c_{\bar{k}}c_{k}\,,\label{H1}
\end{eqnarray}
where $\varepsilon_k$ are the single particle energies (the same for $k$ and $\bar{k}$
states) and the pairing interaction moves pairs of fermions from $k,\bar{k}$ to
$k',\bar{k}'$. We will focus on an equally spaced single particle spectrum
$\varepsilon_{k+1}-\varepsilon_k=\varepsilon$ $\forall$ $k$,  with  a constant
coupling strength $G_{kk'}=G\geq 0$ $\forall$ $k,k'$, and examine the
half-filled case where the number of fermions is $N=\Omega$, with $\Omega$ {\it
even}.

The exact ground state will then be a linear combination of Slater Determinants with
fixed fermion number $N$ and fully occupied or empty pairs $(k,\bar{k})$:
\begin{equation}|\Psi\rangle=\sum_\nu \alpha_\nu
[ \prod_{k}(c^\dagger_kc^\dagger_{\bar{k}})^{n_k^\nu}]|0\rangle\,,\label{GSx}
\end{equation}
where $n_k^\nu=0,1$ indicates the occupation of pair $k,\bar{k}$ ($\sum_k n^\nu_k=N/2$) and   
$\nu=1,\ldots,\binom{\Omega}{N/2}$ runs over these Slater Determinants 
($\sum_\nu|\alpha^2_{\nu}|=1$). This state leads to a single particle density matrix which
remains strictly diagonal in the unperturbed single particle basis: $\langle
c^\dagger_{k}c_{\bar{k}'}\rangle=0\,\forall\,k,k'$ and
\begin{equation}
\langle c^\dagger_kc_{k'}\rangle=\langle c^\dagger_{\bar{k}}c_{\bar{k}'}\rangle=\delta_{kk'} f_k\,,
\label{Ek}
\end{equation}
where $f_k=\langle c^\dagger_kc_k\rangle=\langle c^\dagger_{\bar{k}}c_{\bar{k}}\rangle=
\sum_\nu |\alpha^2_{\nu}| n_k^\nu$ is the average occupation of single particle state $k$ or $\bar{k}$ in the exact ground state 
(\ref{GSx}) ($2\sum_k f_k=N$). 
Since no off-diagonal terms arise, these $f_k$ are the eigenvalues of $\rho^{\rm sp}$.  

\subsubsection{One-body entanglement entropy and global bipartite entanglement}
The exact one-body entanglement entropy (\ref{1}) then becomes
\begin{equation}E(|\Psi\rangle)=2\sum_k h(f_k)\,,\label{Sk}
\end{equation}
where $h(f_k)=-f_k\log_2 f_k-(1-f_k)\log_2(1-f_k)$ 
 represents the entropy $S(\rho_k)$ of the single mode  density [Eq.\ (\ref{ri})]
\begin{equation}
\rho_{k}=\rho_{\bar{k}}=\begin{pmatrix}f_k &0\\0&1-f_k\end{pmatrix}\,,\label{rk}
\end{equation}
i.e., the single mode entropy. We remark that for the exact ground state, $E(|\Psi\rangle)=E^{\rm qsp}(|\Psi\rangle)$ 
since $|\Psi\rangle$ has a fixed $N$. We also note that  $h(f_k)$ is an increasing function of the occupation number fluctuation 
\begin{equation}
\langle n_k^2\rangle-\langle n_k\rangle^2=f_k(1-f_k)={\textstyle\frac{1}{4}}S_2(\rho_k)\,,\label{fluc}\end{equation} 
where $S_2(\rho)=2{\rm Tr}\,[\rho(\mathbbm{1}-\rho)]$ is the quadratic (also called linear) entropy. Eq.\ (\ref{fluc}) 
is then also a measure of single mode entanglement. The relation between entanglement and fluctuations (and also 
higher order cumulants) has been discussed in detail in \cite{SL.15}.   

\begin{figure}[h]
    \begin{center}
        \includegraphics[scale=0.7]{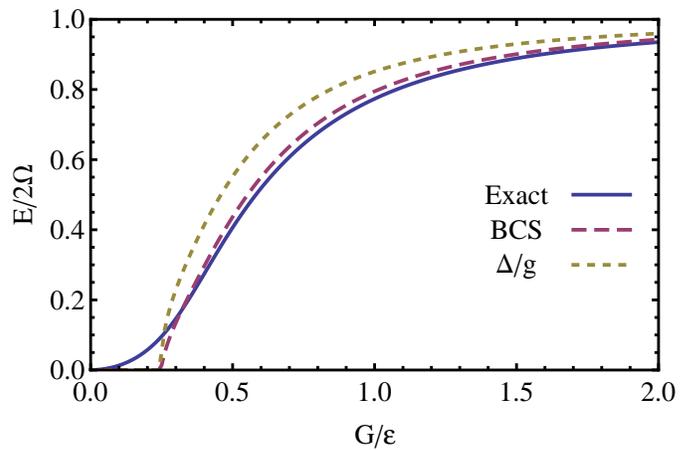}
\caption{Intensive one body entanglement entropy $E(|\Psi\rangle)/(2\Omega)$
(Eq.\ (\ref{Sk}), dimensionless), in the ground state of the  Hamiltonian (\ref{H1}) as a
function of the (dimensionless) scaled coupling strength $G/\varepsilon$ ($\varepsilon$ is  the single particle level spacing) for
$2\Omega=32$ single particle levels and $N=\Omega$. Exact and BCS results are depicted. The
scaled BCS gap $\Delta/g$, with $g=G\Omega/2$,  is also shown. All quantities
approach $1$ for $G/(\Omega\varepsilon)\rightarrow\infty$. BCS results vanish for $G<G_c$. 
Quantities plotted are dimensionless in all figures.} \label{f1}
    \end{center}
    \end{figure}

A plot of (\ref{Sk}) for  the exact ground state (obtained by diagonalization of $H$) of 
a system with $2\Omega=32$ single particle states is depicted in Fig.\  \ref{f1}. This
entropy, which measures the deviation of (\ref{GSx}) from a Slater determinant, is seen here to
be a direct indicator of pairing correlations, becoming nonzero for all $G>0$ and large in the BCS
superconducting phase $G>G_c$ (see \ref{IIIB}). Its behavior for $G>G_c$ resembles in
fact that of the scaled BCS gap $\Delta/G$ (also depicted). Pairing correlations smooth out the original Fermi surface, 
leading to finite average occupations $0<f_k<1/2$ for single particle levels above the Fermi level, which increase with increasing 
$G$, and $1/2<f_k<1$ for levels below the Fermi level, which decrease 
with increasing $G$. Then each term $h(f_k)$ in the sum (\ref{Sk}) increases as  $G$ increases, 
giving rise to the previous behavior of $E(|\Psi\rangle)$. While for $G>G_c$ these effects can be correctly  
described with the BCS approach, in a finite system pairing effects in the exact ground state become also visible within 
the weak coupling sector $0<G<G_c$, where BCS results vanish.  For any  $G>0$ and finite $\Omega$, 
the coupling will mix all  states in the expansion (\ref{GSx}), leading  to $\alpha_{\nu}>0$ $\forall$ $\nu$ 
and hence to $0<f_k<1$ $\forall$ $k$. The state (\ref{GSx})  will then cease to be a strict Slater determinant  as soon as $G$ 
increases from $0$ (see also end of Appendix \ref{C}).

\begin{figure}[h]
    \begin{center}
        \includegraphics[scale=0.7]{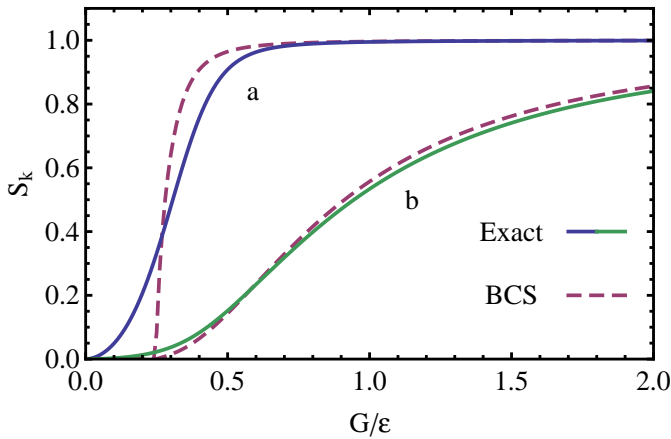}
\caption{The entanglement entropy $S_k=S(\rho_k)=h(f_k)$ of a single particle  mode $k$ with
the rest of the system, for a level closest to the Fermi level ($k=\Omega/2$)
(a) and most distant from the Fermi level ($k=1$) (b), in the system of Fig.\
\ref{f1}. Exact and BCS results are depicted.}
        \label{f2}
    \end{center}
\end{figure}

As seen in Fig.\ \ref{f2}, the increase with $G$ of the single mode entropies $h(f_k)$ will obviously be 
initially more rapid for levels close to the Fermi level,  since their occupation will be more strongly 
affected by the coupling. The occupation number fluctuation $f_k(1-f_k)$ rapidly increases for these levels, 
leading to a larger $h(f_k)$. The finite value of  $E(|\Psi\rangle)$ for $0<G<G_c$ is precisely due to the 
non-negligible contributions $h(f_k)$ from levels close to the Fermi-level (curve (a) in Fig.\ \ref{f2}).  
Nonetheless, for sufficiently large $G$ all levels reached by the coupling become significantly affected. 

In the strong pairing limit  $G\gg \Omega\varepsilon$, $E(|\Psi\rangle)$ and
all $h(f_k)$ saturate for $N=\Omega$, reaching their upper bounds: In this
limit each term  in the sum (\ref{GSx}) will have the same weight, implying,
for $N=\Omega$,
\begin{equation}\alpha_\nu\mathop{\rightarrow}_{G/(\Omega\varepsilon)\rightarrow\infty}
\frac{1}{\sqrt{\binom{\Omega}{\Omega/2}}}\,.\label{as}
\end{equation}
Eq.\ (\ref{as}) leads to $f_k\rightarrow 1/2$ and hence to $h(f_k)\rightarrow
1$ $\forall$ $k$ (entailing maximum fluctuation $f_k(1-f_k)\rightarrow 1/4$), 
implying $E(|\Psi\rangle)\rightarrow 2\Omega$.

\begin{figure}[h]
        \begin{center}
            \includegraphics[scale=0.7]{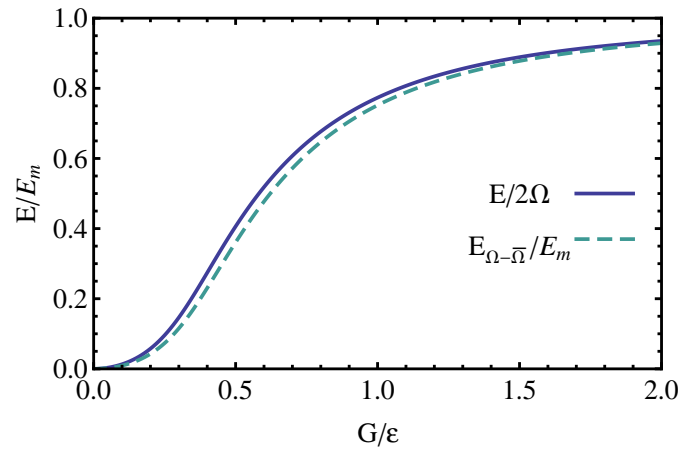}
    \caption{The exact intensive one-body entanglement entropy (solid line) together with the bipartite 
    	    $\Omega-\bar{\Omega}$ entanglement entropy (\ref{SE}) between all modes $k$ and their time-reversed partners $\bar{k}$  
        (dashed line), scaled to its maximum value,   in the system of Fig.\ \ref{f1}. In BCS these two quantities are identical.}
            \label{f3}
        \end{center}
    \end{figure}

The entanglement generated by the pairing correlations can also be seen at the
bipartite level, by considering the bipartition of the full single particle space formed by
the $\Omega$ single particle states $k$ and the $\Omega$ single particle states $\bar{k}$ (${\cal
H}={\cal H}_\Omega\oplus{\cal H}_{\bar{\Omega}}$). For such partition the
expression (\ref{GSx}) is already the {\it Schmidt decomposition} of $|\Psi\rangle$,
since each term in the sum involves orthogonal Slater Determinants at each side. The
associated entanglement entropy is then
\begin{equation}
E_{\Omega-\bar{\Omega}}(|\Psi\rangle)=-\sum_{\nu}|\alpha^2_\nu|\log_2|\alpha^2_\nu|\,.
\label{SE}\end{equation} 
At the BCS level, this entropy is, remarkably, just half the one-body entropy
(\ref{Sk})  (see Eq.\ (\ref{SEB})). In the exact result, this relation holds
approximately for finite $\Omega$. Eq.\ (\ref{SE}) also increases  with increasing $G/\varepsilon$ and 
saturates for $G/(\Omega\varepsilon)\rightarrow\infty$, where it reaches its
upper bound compatible with a fixed $N$: $E_{\Omega-\bar{\Omega}}^{\rm
max}=\log_2\binom{\Omega}{\Omega/2}$ for $N=\Omega$. When scaled to its maximum value,  
$E_{\Omega-\bar{\Omega}}/E_{\Omega-\bar{\Omega}}^{\rm max}$   lies in fact very close to
$E(|\Psi\rangle)/(2\Omega)$, as seen in Fig.\ \ref{f3}. Note also that for large $\Omega$, 
i.e.\ for a system with a large number $N=\Omega$ of particles,  $\log_2\binom{\Omega}{\Omega/2}\approx \Omega$ at leading order, 
which is half the saturation value of $E(|\Psi\rangle)$. This entails $E_{\Omega-\bar{\Omega}}=\frac{1}{2}E(|\Psi\rangle)$ 
in this limit, as in BCS.

\subsubsection{Entanglement of reduced states}
Regarding now the reduced state $\rho_{k\bar{k}}$ of a pair of modes
$(k,\bar{k})$, just the outer $2\times 2$ even parity block in Eq.\
(\ref{rij}), involving here the states $|0\rangle$ and $c^\dagger_k
c^\dagger_{\bar{k}}|0\rangle$, will be nonzero,   since the exact ground state contains
no broken pairs and hence $\langle c^\dagger_k c_k
c_{\bar{k}}c^\dagger_{\bar{k}}\rangle=0=\langle
c^\dagger_{\bar{k}}c_k\rangle=\langle c_kc_{\bar{k}}\rangle$. Since $\langle
c^\dagger_kc^\dagger_{\bar{k}}c_{\bar{k}}c_k\rangle=f_k$, this block will then
be identical to (\ref{rk}), implying, $\forall$ $k$,
   \begin{equation}S(\rho_{k\bar{k}})=S(\rho_k)=S(\rho_{\bar{k}})=h(f_k)\,.\label{srk}\end{equation}
Thus, there is a classical-like correlation between single particle modes $k$ and $\bar{k}$,  captured by the mutual information
\begin{equation}
I_{k\bar{k}}=S(\rho_k)+S(\rho_{\bar{k}})-S(\rho_{k\bar{k}})=h(f_k)\,,\label{irk}
\end{equation}
which is then identical to the single mode entropy. Nonetheless, 
there is no entanglement between them   since $\rho_{k\bar{k}}$ is
diagonal in the basis $\{c^\dagger_kc^\dagger_{\bar{k}}|0\rangle,\,|0\rangle\}$
(this also shows that $E^{\rm qsp}(\rho_{k\bar{k}})=E(\rho_{k\bar{k}})=0$, as
previously stated).

We can also omit states with broken pairs in the reduced density matrix of four
single particle modes $(k\bar{k},k'\bar{k}')$, $k\neq k'$. The ($16\times 16$) matrix
$\rho_{k\bar{k}k'\bar{k}'}$  then reduces to an effective $4\times 4$ non-zero
block $\rho^r_{k\bar{k}k'\bar{k}'}$, with support on the even number parity states
$\{c^\dagger_{k}c^\dagger_{\bar{k}}
c^\dagger_{k'}c^\dagger_{\bar{k}'}|0\rangle$, $c^\dagger_k
c^\dagger_{\bar{k}}|0\rangle$, $c^\dagger_{k'} c^\dagger_{\bar{k}'}|0\rangle$,
$|0\rangle\}$:
\begin{equation}
\rho^r_{k\bar{k}k'\bar{k}'}=\begin{pmatrix}\langle n_{k\bar{k}}
n_{k'\bar{k}'}\rangle&0&0&0\\0&\langle n_{k\bar{k}} \tilde{n}_{k'\bar{k}'}\rangle&
\langle c^\dagger_{k}c^\dagger_{\bar{k}}c_{\bar{k}'}c_{k'}\rangle&0\\
0&\langle c^\dagger_{k'}c^\dagger_{\bar{k}'}c_{\bar{k}}c_{k}\rangle&\langle \tilde{n}_{k\bar{k}} n_{k'\bar{k}'}\rangle&0\\0&0&0&
\langle \tilde{n}_{k\bar{k}} \tilde{n}_{k'\bar{k}'}\rangle\end{pmatrix}\label{rkkp}\end{equation}
Here $n_{k\bar{k}}= c^\dagger_k c_k c^\dagger_{\bar{k}}c_{\bar{k}}$,
$\tilde{n}_{k\bar{k}}=c_kc^\dagger_k c_{\bar{k}}c^\dagger_{\bar{k}}$ are non-zero iff
the pair $(k,\bar{k})$ is fully occupied  or empty respectively. In contrast with $\rho_{k\bar{k}}$,
$\rho_{k\bar{k}k'\bar{k}'}$ will contain quantum correlations due to the nonzero off-diagonal element.

Its fermionic concurrence $C_{kk'}\equiv C(\rho_{k\bar{k}k'\bar{k}'})$, which
determines the  entanglement of formation $E(\rho_{k\bar{k}k'\bar{k}'})=E^{\rm
qsp}(\rho_{k\bar{k}k'\bar{k}'})$ through Eq.\ (\ref{Ef3}), becomes, using  Eq.\
(\ref{Cf}),
\begin{equation}
C_{kk'}=2{\rm Max}[|\langle c^\dagger_{k}c^\dagger_{\bar{k}}c_{\bar{k}'}c_{k'}\rangle|-
\sqrt{\langle n_{k\bar{k}} n_{k'\bar{k}'}\rangle\langle
\tilde{n}_{k\bar{k}}\tilde{n}_{k'\bar{k}'}\rangle},0]\,,\label{Ckkp}
\end{equation}
which will be non-zero for $G>0$. Eq.\ (\ref{Ckkp}) also  represents here the
{\it bipartite} concurrence \cite{W.98} between modes  $k\bar{k}$ and $k'\bar{k}'$, where
each ``side'' is analogous to a qubit (it  can be either empty or fully
occupied in the ground state), forming together a two-qubit system (see Appendix
\ref{C}). The associated bipartite entanglement of formation is just
$E_{kk'}=E^{\rm qsp}(\rho_{k\bar{k}k'\bar{k}'})/4$. Thus, the fermionic
entanglement of $\rho_{k\bar{k}k'\bar{k}'}$ can  be directly identified here
with a {\it bipartite} entanglement. It is also verified, by simple use of Wick's
theorem,  that $C_{kk'}$ vanishes identically  in the BCS approximation (see
next section), so that this entanglement lies strictly {\it beyond} the
standard BCS description, in contrast with previous quantities. As stated before, 
$C_{kk'}$ vanishes for all fermionic gaussian states, which include in particular the BCS ground state. A finite 
concurrence requires sufficiently strong two-body correlations. 

\begin{figure}[h]
    \begin{center}
        \includegraphics[scale=.7]{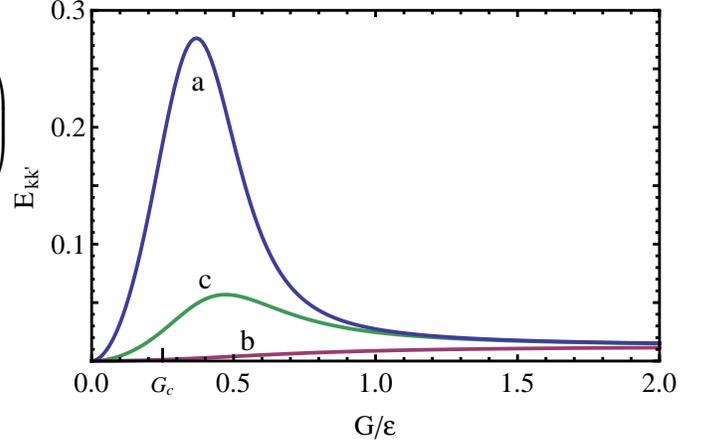}
\caption{Entanglement of formation $E_{kk'}$ determined by the concurrence
(\ref{Ckkp}) between  the modes  $k\bar{k}$ and $k'\bar{k}'$, for pairs closest
($k=\Omega/2$, $k'=k+1$) (a) and most distant ($k=1,k'=\Omega$) (b) to the
Fermi level, and also for pairs next to closest ($k=\Omega/2-1$, $k'=k+3$)
(c), in the system of Fig.\ \ref{f1}.  The BCS result for this quantity
vanishes identically. The peak in (a) occurs close to the BCS superconducting
transition.}
        \label{f4}
    \end{center}
\end{figure}

As seen in  Fig.\ \ref{f4}, $E_{kk'}$ exhibits, remarkably,  a  prominent peak in the vicinity of 
the BCS superconducting transition region $G\approx G_c$ for the pair of
levels closest to the Fermi level, becoming then small for $G\gg G_c$. This
peak is obviously also present in the concurrence $C_{kk'}$ ($E_{kk'}$ is just
an increasing function of $C_{kk'}$), and its height rapidly decays for  levels
$k,k'$ not too close to the Fermi surface,  disappearing for distant levels.

The previous behavior can be understood by analyzing first the strong superconducting 
limit $G/(\Omega\varepsilon)\rightarrow\infty$,
where $\rho^r_{k\bar{k}k'\bar{k}'}$ will be independent of $k, k'$, according to Eq.\
(\ref{as}): The diagonal terms
$\langle n_{k\bar{k}} n_{k'\bar{k}'}\rangle$ and $\langle
\tilde{n}_{k\bar{k}}\tilde{n}_{k'\bar{k}'}\rangle$ in (\ref{rkkp}), which are
the probabilities of finding both pairs fully occupied or empty  become, for
$N=\Omega/2$,
\begin{equation}{\textstyle
\langle n_{k\bar{k}} n_{k'\bar{k}'}\rangle=\langle \tilde{n}_{k\bar{k}}\tilde{n}_{k'\bar{k}'}\rangle=
\frac{\binom{\Omega-2}{\Omega/2}}{\binom{\Omega}{\Omega/2}}=\frac{\Omega-2}{4(\Omega-1)}}\,,
\label{as2}\end{equation}
 while  all elements of the inner block  become equal,
\begin{equation}{\textstyle\langle n_{k\bar{k}} \tilde{n}_{k'\bar{k}'}\rangle
=\langle \tilde{n}_{k\bar{k}} n_{k'\bar{k}'}\rangle
    =\langle c^\dagger_kc^\dagger_{\bar{k}}c_{\bar{k}'} c_{k'}\rangle=
\frac{\binom{\Omega-2}{\Omega/2-1}}{\binom{\Omega}{\Omega/2}}=
\frac{\Omega}{4(\Omega-1)}}\,,\label{as3}\end{equation}
implying it will have a {\it single} non-zero eigenvalue $\frac{\Omega}{2(\Omega-1)}$.

With Eqs.\ (\ref{as2})--(\ref{as3}),  Eq.\ (\ref{Ckkp}) leads in this limit to
\begin{equation}C_{kk'}=\frac{1}{\Omega-1}\,,\label{as5}\end{equation}
$\forall$ $k\neq k'$, decreasing as $\Omega^{-1}$ for large $\Omega$ (i.e, for a  large number $N=\Omega$ of fermions) 
and implying  $E_{kk'}\approx\frac{1}{2}\Omega^{-2}\log_2(2\Omega\sqrt{e})$. 
Eq.\ (\ref{as5})  is in 
agreement with the result for fully connected systems \cite{MRC.08} and the monogamy property of the concurrence 
\cite{CKW.61,OB.06} ($\sum_{k'\neq k}C^2_{kk'}\leq C^2_{k,\{k'\neq k\}}$). All $\rho^r_{k\bar{k}k'\bar{k'}}$ become  
equally entangled in this limit for $k\neq k'$, implying that $C_{kk'}$ should scale 
with $\Omega^{-1}$ for large $\Omega$. 

Thus, for not too small $\Omega$, monogamy prevents a significant entanglement between pairs $k\bar{k}$ and
$k'\bar{k}'$ in the strong superconducting regime. In contrast, at the onset of superconductivity ($G\approx G_c$) 
just the levels $k,k'$ closest to the Fermi level are affected by the pairing correlations, originating the initial 
increase and ensuing peak in the corresponding $C_{kk'}$ and $E_{kk'}$. As $G$ increases further, $C_{kk'}$ becomes 
appreciable for an increasing number of pairs $k\neq k'$ around the Fermi level and the highest $C_{kk'}$ must 
then decrease to comply with monogamy requirements. For large $\Omega$, $C_{kk'}$ is then significant only for $k,k'$ 
close to the Fermi level and just at the transition region, rather than at the strong superfluid regime.

 \begin{figure}[h]
    \begin{center}
        \includegraphics[scale=0.7]{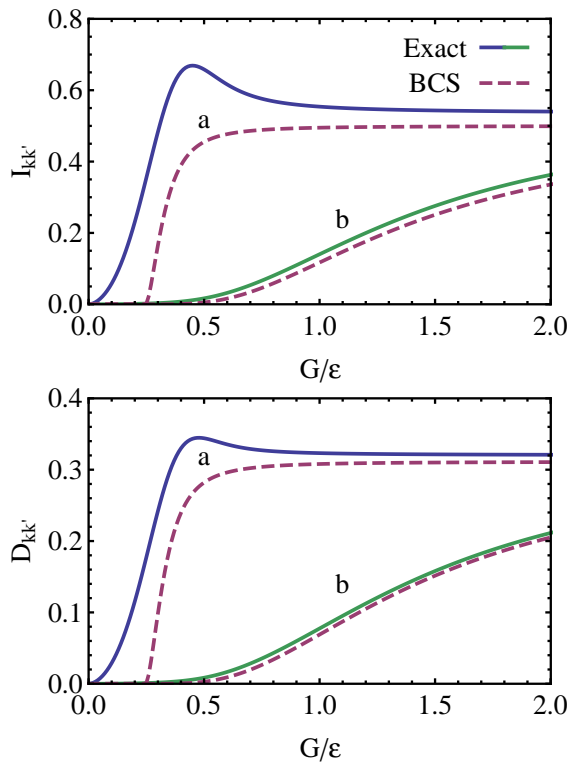}\\
\caption{The mutual information (\ref{Ikkp}) (top) and the quantum discord
(bottom)  between modes $k\bar{k}$ and $k'\bar{k}'$, for pairs closest (a) and
most distant (b) to the Fermi level, in the system of Fig.\ \ref{f1}. Both
quantities are significant in the superconducting phase. The BCS estimate is
now non-vanishing for $G>G_c$.}
        \label{f5}
    \end{center}
\end{figure}
In the latter, correlations between modes $k\bar{k}$ and $k'\bar{k}'$ in  the
reduced state $\rho_{k\bar{k}k'\bar{k}'}$ do exist, but  lead just to a finite
mutual information  $I_{kk'}$ and finite {\it quantum discord} $D_{kk'}$ (see
appendix \ref{C}), as shown in Fig.\ \ref{f5}. In the limit
$G/(\Omega\varepsilon)\rightarrow\infty$,  Eqs.\ (\ref{as2})--(\ref{as3}) lead
to $S(\rho^r_{k\bar{k}k'\bar{k}'}) \approx\frac{1}{2}(3-\Omega^{-1})$ for
 large $\Omega$, implying that the mutual information
 \begin{eqnarray}
 I_{kk'}&\equiv&S(\rho_{k\bar{k}})+S(\rho_{k'\bar{k}'})-S(\rho_{k\bar{k}k'\bar{k}'}) \nonumber\\
 &=&h(f_k)+h(f_{k'})-S(\rho^r_{k\bar{k}k'\bar{k}'})\,, \label{Ikkp}
 \end{eqnarray}
approaches in this limit a finite common value $\approx \frac{1}{2}(1+\Omega^{-1})$ for all $k\neq k'$. 
Both $I_{kk'}$ and $D_{kk'}$  are initially obviously larger for pairs $k,k'$ close to the fermi level, 
in which case they develop a moderate peak at the transition region, but
remain significant for $G\gg G_c$ since they are not restricted by a monogamy property \cite{RCC.10}.  
And in contrast with $E_{kk'}$,  they can be
correctly estimated through BCS. The finite value of the discord (whose
calculation details and asymptotic expression are discussed in Appendix 
\ref{C}) indicates that in the strong pairing regime, the correlations between
pairs $k\bar{k}$ and $k'\bar{k}'$, while not leading to a significant
entanglement of the reduced state, are not fully classical either. As seen from
Eqs.\ (\ref{rkkp})--(\ref{as3}), $\rho^r_{k\bar{k},k'\bar{k}'}$ does not
become diagonal in a product basis, having instead a maximally entangled
non-degenerate eigenstate in the inner block, which leads to the previous
non-classical effect (finite discord). 

In the smallest non-trivial case $\Omega=2$,  $\rho_{k\bar{k}k'\bar{k}'}=|\Psi\rangle\langle\Psi|$ 
becomes pure ($k=1,k'=2$). Hence, the discord coincides with $E_{kk'}$, 
which in turn becomes proportional the one-body entanglement entropy and also the entropy (\ref{SE}):  
$D_{kk'}=E_{kk'}=E_{\Omega-\bar{\Omega}}=E(|\Psi\rangle)/4=h(f_k)=h(f_k')$, where $f_k$ and $f_{k'}=1-f_k$ 
are the (two-fold degenerate) eigenvalues 
of $\rho^{\rm sp}$ (see end of Appendix \ref{C}). All previous measures  then collapse to a single value. 

\subsection{The BCS description\label{IIIB}}
\subsubsection{Standard treatment}
As is well known, the BCS approximation to the ground state of Hamiltonian (\ref{H1}) relies on a state of the form \cite{BCS,RS.80}
\begin{equation}
|\text{\small BCS}\rangle=[\prod_k(u_k+v_k c^\dagger_k c^\dagger_{\bar{k}})]|0\rangle,\label{BCS}
\end{equation}
for even $N$, where $|u_k^2|+|v_k^2|=1$. Such state is the vacuum of quasiparticle fermion operators 
\begin{equation}
a_k=u_kc_k-v_k c^\dagger_{\bar{k}}, \;\;a_{\bar{k}}=v_k c^\dagger_k+u_k c_{\bar{k}}\,,
\end{equation}
satisfying $a_k|\text{\small BCS}\rangle=a_{\bar{k}}|\text{\small
BCS}\rangle=0$ together with fermion anticommutation relations. The
coefficients minimizing $\langle H\rangle=\langle\text{\small
BCS}|H|\text{\small  BCS}\rangle$ under a fixed $\langle N\rangle=2\sum_k
|v_k^2|$ constraint can be chosen real nonnegative, and are given by
$^{u_k}_{v_k}=\sqrt{\frac{\lambda_k\pm\tilde{\varepsilon}_k}{2\lambda_k}}$,
where  $\lambda_k=\sqrt{\tilde{\varepsilon}_k^2+\Delta^2}$ are the
quasiparticle energies, $\tilde{\varepsilon}_k=\varepsilon_k-\mu$, with $\mu$
the chemical potential and $\Delta=G\sum_{k}\langle c_{\bar{k}}c_{k}\rangle$ is
the BCS gap (we have dismissed minor effects in $\tilde{\varepsilon}_k$
stemming from self-energy terms). For $\langle N\rangle=\Omega$ and a uniformly
spaced spectrum, $\mu=\frac{1}{\Omega}\sum_k\varepsilon_k$.

As $\langle c_{\bar{k}}c_k\rangle=u_kv_k=\Delta/(2\lambda_k)$, $\Delta$ is determined by 
the gap equation 
\begin{equation}\Delta=G\Delta\sum_k\frac{1}{2\lambda_k}\,.\label{gap}\end{equation}
The superconducting phase corresponds to the non-trivial solution $\Delta>0$,
which arises for $G>G_c$ with
\begin{equation}
G_c=\frac{2}{\sum_k\frac{1}{|\tilde{\varepsilon}_k|}}\approx \frac{\varepsilon}{\ln(\Omega/2)+\gamma}
\,,\label{gc}
\end{equation}
where the last expression holds  for large $\Omega$ in the equally spaced case
($\gamma=-\frac{\Gamma'[1/2]}{\Gamma[1/2]}\approx 1.96$). For $G<G_c$, $\Delta=0$.

While $E^{\rm qsp}(|\text{\small BCS}\rangle)=0$ $\forall$ $G$, as
$|\text{\small BCS}\rangle$ is a quasiparticle vacuum, the one-body
entanglement entropy
\begin{equation}E(|\text{\small BCS}\rangle)={\rm tr}\,[h(\rho^{\rm sp}_{\rm BCS})]=2\sum_k h(f_k)\,,
\label{Ebcs}\end{equation}
is finite  for $G>G_c$ and provides an excellent approximation to the exact
$E(|\Psi\rangle)=E^{\rm qsp}(|\Psi\rangle)$ in the superconducting phase, as
seen in Fig.\ \ref{f1}. Here $\rho^{\rm sp}_{\rm BCS}=\langle\text{\small
BCS}|\mathbbm{1}-\bm{c}\bm{c}^\dagger|\text{\small BCS}\rangle$ is the BCS single particle 
density matrix, which is diagonal in the unperturbed single particle basis, with eigenvalues
 $f_k=\langle c^\dagger_k c_k\rangle=\langle  c^\dagger_{\bar{k}}c_{\bar{k}}\rangle$ given by
\begin{equation}
f_k=|v_k^2|=\frac{1}{2}(1-\frac{\tilde{\varepsilon}_k}{\lambda_k})\,.
\label{fkbcs}\end{equation} For $G<G_c$, $v_k^2=1$ $(0)$ for levels below
(above) the Fermi level and (\ref{Ebcs}) vanishes, whereas for $G>G_c$,
$v_k^2\in(0,1)$, smoothing the Fermi surface and leading to a finite value of
(\ref{Ebcs}). This indicates the departure of (\ref{BCS}) from a standard Slater  determinant.
For $G/(\Omega\varepsilon)\rightarrow\infty$ and $N=\Omega$, Eq.\ (\ref{gap})
leads to $\Delta\approx G\Omega/2$ and $v_k^2\approx
\frac{1}{2}(1-\tilde{\varepsilon}_k/\Delta)$, implying $E(|\text{\small
BCS}\rangle)\approx 2\Omega[1-\frac{\sum_k
\tilde{\varepsilon}_k^2}{2\Omega\Delta^2\ln 2}]$, which saturates in the limit.

Eq.\ (\ref{fkbcs}) also provides a good approximation to the single mode
entropies $S(\rho_k)=h(f_k)$, as seen in Fig.\ \ref{f2}. 
As stated before, each term $h(f_k)$ is an increasing function of 
the occupation number fluctuation $f_k(1-f_k)$, which in BCS becomes $u_k^2v_k^2$. And 
the quadratic one-body entanglement entropy $S_2(\rho^{\rm sp})=
4\,{\rm tr}\,[\rho^{\rm sp}(\mathbbm{1}-\rho^{\rm sp})]=8\sum_k u_k^2v_k^2$ 
is in BCS just twice the total number fluctuation $\langle N^2\rangle-\langle N\rangle^2$ \cite{RS.80}. 
Of course, this relation is not valid in the exact ground state, for which the number fluctuation is strictly zero.   

The BCS state (\ref{BCS})  does not have a fixed fermion number but has a
definite (positive) number parity. It is in fact of the same form (\ref{GSx}) with
$\alpha_{\nu}=\prod_k v_k^{n_k^\nu}\,u_k^{1-n_k^\nu}$  and $n_k^\nu=0,1$ 
{\it independent} variables, such that $\nu=1,\ldots,2^\Omega$. 
Thus, the entanglement entropy (\ref{SE}) between
all modes $k$ and all modes $\bar{k}$ reduces to the entropy of a product of
$\Omega$ independent density operators with eigenvalues $|v_k^2|=f_k$ and
$|u_k^2|=1-f_k$. Hence,
\begin{equation}E_{\Omega-\bar{\Omega}}(|\text{\small BCS}\rangle)=
\sum_k h(f_k)=\frac{1}{2}E(|\text{\small BCS}\rangle)\,.\label{SEB}\end{equation}
A similar relation holds approximately  in the exact result (Fig.\ \ref{f3}). An entropy similar 
to (\ref{SEB}) was defined in \cite{Pu.14} for the BCS state and analyzed in the continuous limit. 

Considering now the reduced state of levels $(k\bar{k})$, BCS leads (using
Wick's theorem \cite{RS.80}) to $\langle c^\dagger_k c_k
c_{\bar{k}}c^\dagger_{\bar{k}}\rangle= \langle c^\dagger_k c_k\rangle\langle
c_{\bar{k}}c^\dagger_{\bar{k}}\rangle-\langle c^\dagger_{k} c^\dagger_{\bar{k}}
\rangle\langle c_{\bar{k}} c_{k}\rangle = v_k^2u_k^2-(u_kv_k)^2=0$ and $\langle
c^\dagger_k c^\dagger_{\bar{k}}c_{\bar{k}}c_{k}\rangle=v_k^4+(u_kv_k)^2=f_k$,
as in the exact case. Hence, Eqs.\ (\ref{srk})--(\ref{irk}) remain valid in BCS
with $f_k$ given by (\ref{fkbcs}).

Differences arise, however, in the four level  density matrix (\ref{rkkp}),
since Wick's theorem implies that all quantities will be a function of the
$f_k$. For $k\neq k'$ we have $\langle n_{k\bar{k}}
n_{k'\bar{k}'}\rangle=\langle n_{k\bar{k}}\rangle\langle
n_{k'\bar{k}'}\rangle$,
 $\langle n_{k\bar{k}}\tilde{n}_{k'\bar{k}'}\rangle=\langle n_{k\bar{k}}\rangle\langle\tilde{n}_{k'\bar{k}'}\rangle$ and
 $\langle \tilde{n}_{k\bar{k}}\tilde{n}_{k'\bar{k}'}\rangle=
 \langle\tilde{n}_{k\bar{k}}\rangle\langle \tilde{n}_{k'\bar{k}'}\rangle$,
 with $\langle n_{k\bar{k}}\rangle=v_k^2=f_k$, $\langle \tilde{n}_{k}\rangle=u_k^2=1-f_k\equiv\tilde{f}_k$ and
 $\langle c^\dagger_{k}c^\dagger_{\bar{k}}c_{\bar{k}'}c_{k'}\rangle=u_kv_k u_{k'}v_{k'}$.
 Hence, in BCS Eq.\ (\ref{rkkp}) becomes
\begin{equation}
\rho^{r\,{\rm BCS}}_{k\bar{k}k'\bar{k}'}=\begin{pmatrix}f_k
f_{k'}&0&0&0\\0&f_k\tilde{f}_{k'}&\sqrt{f_k\tilde{f}_kf_{k'}\tilde{f}_{k'}}&0\\
0&\sqrt{f_k\tilde{f}_kf_{k'}\tilde{f}_{k'}}&\tilde{f}_kf_{k'}
&0\\0&0&0&\tilde{f}_k\tilde{f}_{k'}\end{pmatrix}\,.\label{rkkpbcs}
\end{equation}
It then  has always just three non-zero eigenvalues ($f_k f_{k'}$,
$\tilde{f}_k\tilde{f}_{k'}$ and $f_k\tilde{f}_{k'}+\tilde{f}_kf_{k'}$), which
in the exact state occurs exactly only for
$G/(\Omega\varepsilon)\rightarrow\infty$. These expressions lead to  $|\langle
c^\dagger_{k}c^\dagger_{\bar{k}}c_{\bar{k}'}c_{k'}\rangle|=\sqrt{\langle
n_{k\bar{k}} n_{k'\bar{k}'}\rangle \langle
\tilde{n}_{k\bar{k}}\tilde{n}_{k'\bar{k}'}\rangle}$ and hence to  $C_{kk'}=0$ 
$\forall$ $\Delta$ and $k\neq k'$, as previously stated. BCS cannot reproduce the concurrence (\ref{Ckkp}) 
since the latter vanishes in any gaussian state, and hence for any $|\Psi\rangle$ which is a Slater determinant or quasiparticle vacuum, 
like the BCS state (\ref{BCS}). A finite concurrence requires sufficiently strong two-body correlations, 
in order to have a positive difference in (\ref{Ckkp}).

Nonetheless, BCS still leads to a good estimate of
$S(\rho_{k\bar{k}k'\bar{k}'})$ and of both  the mutual information $I_{kk'}$ and quantum discord $D_{kk'}$ in the 
superconducting phase $G>G_c$, as seen in Fig.\ \ref{f5}. Moreover, for
$G/(\Omega\varepsilon)\rightarrow\infty$,  $f_k\rightarrow 1/2$ $\forall$ $k$,
and the exact  limits $I_{kk'}=1/2$, $S(\rho_{kk'})=3/2$ and
$D_{kk'}=\frac{3}{2}-3\frac{\log_2 3}{4}$ (see Appendix \ref{C}) are obtained for large $\Omega$.

\subsubsection{Number projected treatment}
One could now ask how the finite value of the concurrence $C_{kk'}$ 
can be predicted in a BCS-based scheme. The answer lies, of course,
in the number projected BCS approximation \cite{RS.80}, based on the state
\begin{equation}P_N|\text{\small BCS}\rangle\propto \sum_{\nu}\left[
\prod_k v_k^{n_k^\nu}\,u_k^{1-n_k^\nu}
    (c^\dagger_k c^\dagger_{\bar{k}})^{n_k^\nu}\right]|0\rangle
\label{PBCS}\,,
\end{equation}
where $P_N=\frac{1}{2\pi}\int_{0}^{2\pi}e^{-i\phi(\hat{N}-N)}d\phi$ is the
projector onto fixed (even) fermion  number $N$,  $n_k^\nu=0,1$ and now $\sum_k
n_k^\nu=N/2$, with $\nu=1,\ldots,\binom{\Omega}{N/2}$ (the $n_k^\nu$ are no longer independent variables). 
The state (\ref{PBCS}) has the same form as the exact state (\ref{GSx}), but with specified coefficients $\alpha_{\nu}=\prod_k
v_k^{n_k^\nu}u_k^{1-n_k^\nu}$.

While projection after variation already improves BCS in the superconducting phase, projection before  
variation can properly describe also the normal sector $G<G_c$, where standard BCS estimates vanish for
all correlation measures. We consider here a simple approach where  the form of
the coefficients $u_k$ and $v_k$ in (\ref{PBCS}) is the same as in standard BCS, but $\Delta$
is left as a variational parameter to be determined from the minimization of
the projected average energy $\langle H\rangle_{N}=\frac{\langle\text{\scriptsize BCS}|P_N H|\text{\scriptsize
BCS}\rangle}{\langle\text{\scriptsize BCS}|P_N |\text{\scriptsize
BCS}\rangle}$. Full self-consistent methods can also be employed \cite{SRLR.02}. 

\begin{figure}[h]
	\begin{center}
		\includegraphics[scale=0.7]{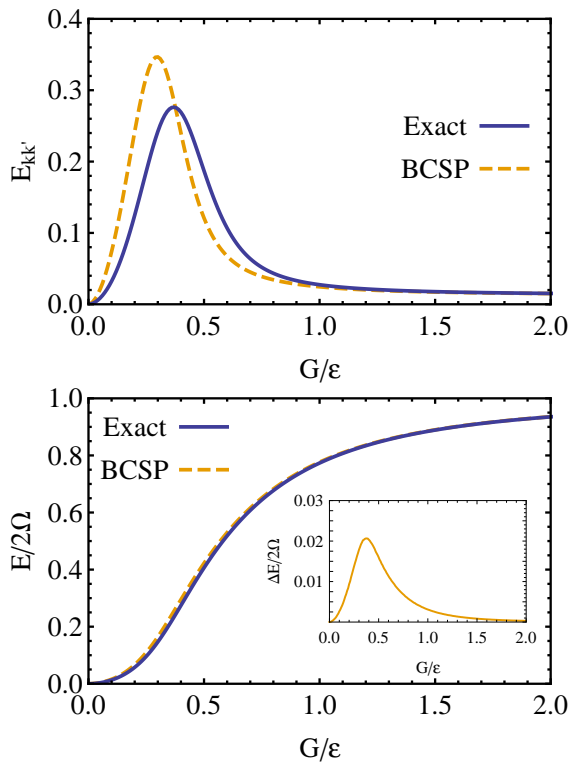}
		\caption{Top: Entanglement of formation $E_{kk'}$ between modes  $k\bar{k}$ and
			$k'\bar{k}'$ for pairs closest to the Fermi level  according to exact and
			projected BCS results, in the system of Fig.\ \ref{f1}. Bottom: The
			corresponding one body entanglement entropy. The projected BCS result is
			accurate for all values of $G$, although the difference $\frac{\Delta
				E}{2\Omega}=\frac{(E_{\rm BCSP} - E_{\rm Exact})}{2\Omega}$, depicted in the
			inset, is also maximum  at the transition region.}
		\label{f6}
	\end{center}
\end{figure}

As seen in Fig.\ \ref{f6}, such an approach  is sufficient to
predict a {\it finite} concurrence $C_{kk'}$, which fairly reproduces the exact
result,  including the peak for pairs  $k\bar{k}, k'\bar{k}'$ close to the
fermi level. The essential reason is that for $\Delta>0$, the projected state (\ref{PBCS}) is no longer gaussian, 
i.e.\ it is not a quasiparticle vacuum nor a Slater determinant, and Wick's theorem no longer holds. 
It  contains  two-body correlations and has in fact a very high overlap with the exact ground state (\ref{GSx}). 

The effective $\Delta$ obtained with projection before variation is positive for {\it all} $G>0$ and exhibits a smooth 
increase with increasing $G$, so that (\ref{PBCS}) will also lead to
quite accurate estimates of all other quantum correlation measures shown in
Figs.\ \ref{f1}-\ref{f3} and \ref{f5}, {\it including} the interval $0<G\leq
G_c$, as seen in the bottom panel of Fig.\ \ref{f6} for the one-body
entanglement entropy. It is also noticed that the transition region $G\approx
G_c$, where $C_{kk'}$ exhibits its peak, is precisely that where the
discrepancy between the exact and the projected BCS predictions is most
significant, as seen in the inset.

\section{Conclusions}
We have analyzed in detail the behavior of general  fermionic entanglement 
measures in the exact ground state of a finite superconducting system. The one-body
entanglement entropy, which represents the minimum distance (as measured by the 
relative entropy) to a fermionic gaussian state, is seen to be here a close 
indicator of pairing correlations, saturating in the strong coupling limit and
behaving like a scaled BCS gap. It is practically proportional to the bipartite
entanglement entropy between all single particle modes $k$ and their time reversed partners $\bar{k}$, being exactly
proportional at the BCS level. BCS  provides in fact a good estimation of
these entropies in the whole superconducting phase.

In contrast, the entanglement of a subset of fermionic modes, determined in
general by a mixed reduced state with no fixed fermion number, can exhibit a
quite different behavior. The first non-trivial case was shown to be that of
four single particle modes $k,\bar{k}, k',\bar{k}'$, whose  entanglement of formation can be
evaluated through the fermionic concurrence and can be  interpreted as a
bipartite mode entanglement. This entanglement vanishes identically in BCS as well as in any fermionic gaussian state. 
In the exact ground state it shows instead a peak in the vicinity of 
 the superconducting transition region for single particle states $k,k'$ close to the Fermi level, 
 which are those most affected by  the coupling at the onset of the transition, possessing then a larger 
 occupation number fluctuation in this region.  The concurrence becomes then small in
the strong coupling regime for not too small $\Omega$ due to monogamy restrictions. Hence, it is here an indicator of 
the transition, reflecting the increased complexity of the exact ground state in this
region. It requires at least a number projected BCS treatment for its
approximate description. We have also shown that while not significantly
entangled, these four modes  do remain correlated in the strong coupling
regime, exhibiting there a finite mutual information and quantum discord, due the non-zero off diagonal 
terms in the density matrix, and showing for this reason a less noticeable peak at the transition region. 
The present results provide then new insights into the relation between fermionic entanglement and
superconducting correlations.

\acknowledgments
We acknowledge support from CIC (RR) and CONICET (MDT, NG), of Argentina, and CONICET Grant PIP 112201501-00732.
Discussions with Prof.\ J.M.\ Matera are also acknowledged. 
\appendix

\section{Minimum relative entropy \label{A}}
Given two density operators $\rho$, $\rho'$ for a given system,
the relative entropy (\ref{srel00}) can be written as \cite{Wh.78,Ve.02}
\begin{equation}S(\rho||\rho')=-{\rm Tr}[\rho\log_2\rho']-S(\rho)\,,
\label{Srel}
\end{equation}
where $S(\rho)=-{\rm Tr}[\rho \log_2\rho]$ is the von Neumann entropy.
It satisfies $S(\rho||\rho')\geq 0$, with $S(\rho||\rho')=0$ iff $\rho=\rho'$ \cite{Wh.78}.
Let us now consider a $\rho'$  of the form
\begin{equation}
\rho'=Z^{-1}\exp[-\sum_{\nu=1}^m \lambda_\nu O_\nu]\,,\label{Op}
\end{equation}
where $Z={\rm Tr}\,\exp[-\sum_\nu\lambda_\nu O_\nu]$ and
$\{O_\nu,\,\nu=1,\ldots,m\}$ is an arbitrary  set of $m$ linearly independent
operators (hermitian or comprising both $O_\nu$ and $O_\nu^\dagger$). This form
of $\rho'$ is that which maximizes the entropy $S(\rho')$ subject to the
constraint of fixed expectation values $\langle O_\nu\rangle$,
$\nu=1,\ldots,m$. It is easy to show that  for fixed $\rho$,
\begin{equation}
\mathop{\rm Min}_{\{\lambda_\nu\}} S(\rho||\rho')=S(\rho')-S(\rho)\label{min}\,,
\end{equation}
with  the minimum  reached for those $\lambda_\nu$ satisfying
\begin{equation}
{\rm Tr}\,[\rho' O_\nu]={\rm Tr}\,[\rho O_\nu],\;\;\nu=1,\ldots,m\,,
\end{equation}
i.e., for that $\rho'$ which reproduces the expectation values
determined by $\rho$  of all operators $O_\nu$ of the chosen set. \\
{\it Proof:} Setting $\langle O_\nu\rangle_\rho\equiv {\rm Tr}\,[\rho O_\nu]$,
we obtain, from Eqs.\ (\ref{Srel})--(\ref{Op}),
\begin{equation}
S(\rho||\rho')=\frac{1}{\ln 2}(\sum_{\nu=1}^k \lambda_\nu \langle O_\nu\rangle_\rho
+\ln Z)-S(\rho)\,.\label{srel}\end{equation}
As $\frac{\partial\ln Z}{\partial \lambda_\nu}=-\langle O_\nu\rangle_{\rho'}$,
equations  $\frac{\partial}{\partial\lambda_\nu}S(\rho||\rho')=0$ lead to
\begin{equation}
\langle O_\nu\rangle_{\rho'}=\langle O_\nu\rangle_\rho\,,\;\;\nu=1,\ldots,m
\end{equation}
in which case Eq.\ (\ref{srel}) reduces to  Eq.\ (\ref{min}). \qed

It then follows that $S(\rho')\geq S(\rho)$, with $S(\rho')=S(\rho)$ iff
$\rho'=\rho$. The minimum relative entropy is then a measure of the information
contained in $\rho$ that cannot be  contained in any operator of the form
(\ref{Op}). If $\rho$ is pure  and the operators $O_\nu$ comprise the full set
of one-body operators $c^\dagger_i c_j$,  (\ref{min}) leads to Eq.\
(\ref{min2}) provided traces are taken in the grand canonical  ensemble.
Similarly, if the $O_\nu$ also include the operators $c_i c_j$ and $c^\dagger_i
c^\dagger_j$, (\ref{min}) leads to Eq.\ (\ref{minqsp})  (again in the full grand canonical 
ensemble).

\section{Fermionic concurrence of four single particle  states \label{B}}
Labeling the four single particle  states  $i,j,k,l$ as $1,2,3,4$, and setting
$|\bar{0}\rangle\equiv c^\dagger_1c^\dagger_2c^\dagger_3c^\dagger_4|0\rangle$,
with $|0\rangle$ the fermionic vacuum, the operator $T$ in $R(\rho_{ijkl})$ in
Eq.\ (\ref{Cf}) is represented, in the basis
$\{|0\rangle,c^\dagger_1c^\dagger_2|0\rangle,
c^\dagger_1c^\dagger_3|0\rangle,c^\dagger_1c^\dagger_4|0\rangle,-|\bar{0}\rangle,
c_2c_1|\bar{0}\rangle,c_3c_1|\bar{0}\rangle,c_4c_1|\bar{0}\rangle\}$, by the
matrix \cite{GR.15}
 \[T=\begin{pmatrix}0&I_4\\I_4&0\end{pmatrix}\,.\]
The same matrix holds for an odd parity state $\rho_{ijkl}$ in the basis
$\{c^\dagger_1|0\rangle,c^\dagger_2|0\rangle,c^\dagger_3|0\rangle,c^\dagger_4|0\rangle$,
 $c_1|\bar{0}\rangle, c_2|\bar{0}\rangle,c_3|\bar{0}\rangle,c_4|\bar{0}\rangle\}$.

\section{Four modes reduced states as two-qubit states and quantum discord \label{C}}

The ground state in (\ref{GSx}) is a superposition of states where pairs of
modes $k\bar k$ are either fully occupied or empty. Following ref.\
\cite{GR.17}, we could think of such pairs as even-parity qubits and use this
representation to see the reduced state (\ref{rkkp}) of the four modes $k{\bar
k},k'{\bar k}'$, as a mixed two-qubit state. From lemma 1 of \cite{GR.17} it
then follows that the fermionic concurrence (\ref{Ckkp}) {\it is} the Wootters
concurrence \cite{W.98} of these two qubits.

Furthermore, fermion operators analogous to the Pauli matrices can be
introduced for these qubits, so that any local operation can be described in
terms of them:
\begin{eqnarray}
\tilde\sigma_k^x&=&c_{k}^\dagger c_{\bar k}^\dagger+c_{\bar k} c_{k}\\
\tilde\sigma_k^y&=&-i(c_{k}^\dagger c_{\bar k}^\dagger-c_{\bar k} c_{k})\\
\tilde\sigma_k^z&=&c_{k}^\dagger c_{k}+c_{\bar k}^\dagger c_{\bar k}-1.
\end{eqnarray}
It is verified that these operators satisfy
$[\tilde\sigma_k^\mu,\tilde\sigma_{k'}^\nu]=2i\delta_{kk'}\epsilon_{\mu\nu\gamma}
\tilde\sigma_k^\gamma$ and
$(\tilde\sigma_k^\mu)^2|\psi\rangle_{k}=|\psi\rangle_{k}$ for any even parity
state  $|\psi\rangle_{k}$ of the pair $k\bar k$. In terms of these operators
any mixed state of these two qubits can be written as
\begin{eqnarray}
\rho_{kk'}&=&\rho_k\rho_{k'}+\frac{1}{4}C_{\mu\nu}\sigma_k^\mu\sigma_{k'}^\nu,\\
\rho_{k}&=&\frac{1}{2}(1+r_{k\mu}\sigma_k^\mu),
\end{eqnarray}
where $r_{k\mu}=\langle\sigma_k^\mu\rangle$ and
$C_{\mu\nu}=\langle\sigma_k^\mu\sigma_{k'}^\nu\rangle-\langle\sigma_k^\mu
\rangle\langle\sigma_{k'}^\nu\rangle$ is the {\it correlation tensor} of the
state. This representation turns out to be convenient to evaluate the quantum
discord  \cite{GR.14}.

Recall that the quantum discord $D(A|B)$ of a state $\rho_{AB}$ of a bipartite
system of distinguishable constituents $A$, $B$ can be defined as the minimum 
difference of two quantum extensions of the conditional entropy
\cite{OZ.01,HV.03,Mo.11}, 
\begin{eqnarray}
D(A|B)&=&\mathop{\rm Min}_{\{\Pi_j\}}S(A|B_{\{\Pi_j\}})-[S(\rho_{AB})-S(\rho_B)]\,,\label{Dab}\\
S(A|B_{\{\Pi_j\}})&=&\sum_j p_jS(\rho_{A/\Pi_j})\label{cent}\,,
\end{eqnarray}
where $\rho_{A(B)}={\rm Tr}_{B(A)}\rho_{AB}$ is the reduced state of subsystem
$A(B)$, the set $\{\Pi_j\}$ describes a local measurement on $B$, $p_j={\rm
Tr}\,[\rho_{AB}\Pi_j]$ is the probability of result $j$ in that measurement and
$\rho_{A/\Pi_j}=p_j^{-1}{\rm Tr}_B[\rho_{AB}\Pi_j]$ the conditional state of $A$
after such result is obtained. Evaluating $D(A|B)$ then requires to find the
minimum over all local measurements of the conditional entropy
$S(A|B_{\{\Pi_j\}})$.

For a two qubit state and for a projective measurement along
direction $\bm{k}$ in the Bloch sphere of qubit $B$, the conditional entropy 
(\ref{cent}) reads, explicitly,
\begin{equation}
S(A|B_{\bm k})=\sum_{\mu=\pm,\nu=\pm}p_{\nu\bm k}f(\lambda^\mu_{\nu\bm k}),\label{centk}
\end{equation}
where $f(x)=-x\log_2 x$, $p_{\nu\bm k}=\frac{1}{2}(1+\nu\bm r_B\cdot\bm k)$ are
the probabilities of the two possible results of such measurement and
\[\lambda^\mu_{\nu\bm k}={\textstyle\frac{1}{2}(1+\mu
 |\bm r_A+\nu\frac{C\bm k}{1+\nu\bm r_B\cdot\bm k}|)}\]
the eigenvalues of the ensuing conditional state $\rho_{A/\nu\bm k}$ of qubit
$A$. It was shown in ref.\ \cite{GR.14} that in the weakly correlated regime,
the measurement minimizing (\ref{centk}) is determined essentially by the
direction of one of the singular vectors of the correlation tensor $C$.

Our interest here is in the state (\ref{rkkp}), which in the present notation
is an $X$-type state symmetric under rotations around the $z$ axis. It has
marginal vectors $\bm r_{k(k')}$ parallel to the singular vector $z_{k(k')}$ of
$C$, i.e., $\bm{r}_k=(0,0,r_{kz})$ with $r_{kz}=2\langle
n_{k\bar{k}}\rangle-1$, and a correlation tensor already diagonal in the chosen
basis, with
\begin{eqnarray}C_{xx}&=&C_{yy}=2\langle c^\dagger_k c^\dagger_{\bar{k}}
c_{\bar{k}'}c_{k'}\rangle, \label{Cxx}\\
C_{zz}&=&4(\langle n_{k\bar{k}}n_{k'\bar{k}'}\rangle-
\langle n_{k\bar{k}}\rangle\langle n_{k'\bar{k}'}\rangle)\,.\label{Czz}
        \end{eqnarray}
Therefore, the minimizing measurement in the weakly correlated limit can be a
projective measurement either along  $z$ or along any vector $\bm{k}$ in the
$xy$ plane. Beyond weak correlation, it is easy to show that for this state the
previous projective measurements are still stationary. Moreover, for reduced
states obtained from the ground state of the present pairing system, we have verified
that the minimum is always obtained for a measurement along any vector $\bm{k}$
in the $xy$ plane, which is precisely that determined by the  pairing
correlations (Eq.\ (\ref{Cxx})).

We then obtain $p_{\nu\bm{k}}=\frac{1}{2}$ and $\lambda^\mu_{\nu \bm
k}=\frac{1}{2}[1+\mu\sqrt{r_{kz}^2+C_{xx}^2}]$ for both $\nu=\pm$. In
particular, in the strong superconducting regime, Eqs.\
(\ref{as2})--(\ref{as3}) lead to $r_{kz}=0$ and
$C_{xx}=\frac{\Omega}{4(\Omega-1)}$, in which  case Eqs.\
(\ref{Dab})--(\ref{centk}) lead to
\begin{equation}
{\textstyle D_{kk'}\approx\frac{1}{2}(1-\frac{\log_2 3} {2})(3+\Omega^{-1})}\,,
\end{equation}
for large $\Omega$. The discord remains then finite in this limit.

{\it The case $N=\Omega=2$}. The case of $N=2$ fermions in 
$\Omega=2$ twofold degenerate levels is the smallest non-trivial pairing
system. The exact ground state of the Hamiltonian (\ref{H1}) for  $G\geq 0$
becomes just $|\Psi\rangle=(\alpha_k
c^\dagger_kc^\dagger_{\bar{k}}+\alpha_{k'}c^\dagger_{k'}c^\dagger_{\bar{k}'})|0\rangle$, 
with $k=1$, $k'=2$, $\alpha_{^k_{k'}}=\sqrt{\frac{\lambda\pm\varepsilon}{2\lambda}}$ and 
$\lambda=\sqrt{\varepsilon^2+G^2}$, which is entangled for $G>0$ (i.e., it is
not a Slater determinant nor a quasiparticle vacuum). The state (\ref{rkkp}) becomes obviously pure, with $\langle
n_{k\bar{k}}n_{k'\bar{k}'}\rangle=\langle
\tilde{n}_{k\bar{k}}\tilde{n}_{k'\bar{k}'}\rangle=0$ and $\langle
n_{k\bar{k}}\tilde{n}_{k'\bar{k}'}\rangle=|\alpha_k^2|$, $\langle
c^\dagger_{k'}c^\dagger_{\bar{k'}}c_{\bar{k}}c_k\rangle=\alpha_k\alpha^*_{k'}=G/(2\lambda)$.
The concurrence (\ref{Ckkp}) reduces to $C=2|\alpha_k\alpha_{k'}|$, i.e., 
\begin{equation}
C_{kk'}=\frac{|G|}{\sqrt{\varepsilon^2+G^2}}\,,
\end{equation}
approaching $1$ for $G/\varepsilon\rightarrow\infty$, in agreement with the
limit (\ref{as5}) for $\Omega=2$. The quantum discord then coincides
exactly with the bipartite entanglement entropy $E_{kk'}$, which here is just 
$E_{\Omega-\bar{\Omega}}$, and is exactly proportional to the one-body entropy 
$E(|\Psi\rangle)=h(\rho^{\rm sp})$: $D_{kk'}=E_{kk'}=S(\rho_{k\bar{k}})=
S(\rho_{k'\bar{k}'})=E_{\Omega-\bar{\Omega}}=E(|\Psi\rangle)/4=-\sum_{k}
|\alpha_k^2|\log_2|\alpha_k^2|=h(f_k)=h(f_{k'})$, with $I_{kk'}=2E_{kk'}$ and 
$f_k=|\alpha_k^2|$, $k=1,2$, the eigenvalues (two-fold degenerate) of $\rho^{\rm sp}$.

\end{document}